# Investigation of steady-state and time-dependent luminescence properties of colloidal InGaP quantum dots

Subhasish Chatterjee[1,3], Nikesh V. Valappil[1] and Vinod M. Menon[1,2] *

[1]Laboratory for Nano and Micro Photonics (LaNMP)-Department of Physics, Queens College-CUNY, Flushing, NY, [2]The Graduate Center of CUNY, New York, NY.
[3]Department of Chemistry, City College of New York and the Graduate Center of CUNY, New York, NY.

**ABSTRACT**

Quantum dots play a promising role in the development of novel optical and biosensing devices. In this study, we investigated steady-state and time-dependent luminescence properties of InGaP/ZnS core/shell colloidal quantum dots in a solution phase at room temperature. The steady state experiments exhibited an emission maximum at 650 nm with full width at half maximum of ~ 85 nm, and strong first-excitonic absorption peak at 600 nm. The time-resolved luminescence measurements depicted a bi-exponential decay profile with lifetimes of $\tau_1$~ 47 ns and $\tau_2$~ 142 ns at the emission maximum. Additionally, luminescence quenching and lifetime reduction due to resonance energy transfer between the quantum dot and an absorber are demonstrated. Our results support the plausibility of using these InGaP quantum dots as an effective alternative to highly toxic conventional Cd or Pb based colloidal quantum dots for biological applications.

**INTRODUCTION**

Quantum dot (QD), often described as an 'artificial atom', exhibits discrete energy levels and the spacing of the energy levels can be precisely modulated through the variation of their size [1]. Consequently, QDs act as robust light emitters with finely tunable fluorescence emission that asserts a great advantage over conventional organic chromophores [2]. Nanoscopic size, stability in organic and aqueous phases, strong fluorescence, and a combination of large molar absorbities and high quantum yields, are the unique properties that make QDs very attractive for biosensing and as fluorescent labels in biological research [2-5]. QDs show promising technological potential in the development of photonic transistors [6], photovoltaic devices [7], light-emitting diodes [8], and lasers [1]. Furthermore, QDs can serve as efficient fluorescent donors in resonance energy transfer (RET) process [2, 9], a powerful tool for structural investigation of biological and synthetic macromolecules [10]. RET phenomenon involving a pair of fluorescent donor and quencher has been extensively utilized to depict biomolecular conformational changes, which are imperative for understanding
the structure-function characteristics of proteins [11] and nucleic acids [12, 13].
The long radiative lifetime of QDs (>10 ns) facilitates continuous and long-term tracking of slow biological process and conformational dynamics of biomolecules with large distance changes, a challenging task with conventional organic fluorophores [2-4]. Indeed, the applications of CdSe and CdS QDs have proven to be promising in this context [3, 4, 9, 12].

In the present study, we have investigated the steady state and time dependent optical properties of InGaP/ZnS core-shell QDs. These QDs can serve as an attractive low toxicity alternative to the widely used semiconductor QDs due to the absence of heavy metal components. Conventional cadmium (Cd) and lead (Pb) based semiconductor nanocrystals produce bio-hazardous wastes, specifically classified under Restriction of Hazardous Substances Directive (RoHS). The use of II-VI and III-V semiconductor QDs (such as, CdTe, CdSe) as fluorescent labels causes a practical challenge in biological applications due to their intrinsic toxicity, owing to the presence of surface ions and the production of photo-initiated radicals [14]. Moreover, the long-term effects of toxic QDs in a biological system are not well understood. The use of InGaP QDs can provide an attractive biocompatible alternative for life science research, such as cell imaging, cell tracking, and cancer assays [2-5]. Since contriving experimental conditions can significantly affect the photo-physical properties of QDs [1,3], a comprehensive understanding of their emission dynamics is essential [15]. Time-dependent analysis of the fluorescence emission of QDs depicts the status of their photo-excited states, which depend on the confinement of electrons in a nearly defect free three-dimensional crystal lattice as well as the surface conditions and local external environment of the QDs [15,16]. In this study, we have investigated the time-resolved photoluminescence (PL) lifetime of InGaP/ZnS core/shell QDs, elucidating the emission dynamics that plays a key role in their application as fluorescent donors [2,4]. In addition, the change in the lifetime of InGaP/ZnS core/shell QDs was demonstrated in the presence of an absorber molecule, validating the applicability of InGaP QDs to RET process.

**EXPERIMENTAL METHODS**

InGaP/ZnS core/shell QDs in toluene were obtained from Evident Technologies Inc. The size of the QD core is ~ 3.5 nm, and the diameter of the InGaP QD including its ZnS shell is approximately 6 nm [17]. The absorption spectrum of InGaP/ZnS core/shell QDs was measured using a Varian (CARY 5000) spectrometer. The continuous wave (CW) PL study of the QD solution was performed using an argon-ion laser as an excitation source with primary excitation wavelength of 488 nm, and the PL spectrum of the QDs was recorded using a fiber-coupled spectrometer (Ocean Optics HR4000).

We used a time-correlated single photon counting (TCSPC) system (Horiba Jobin-Yvon) to investigate the time-dependent optical properties of the QD emission. A 50 ps diode laser operating at 100 kHz repetition rate and 467 nm emission wavelength was used as the excitation source. To investigate the RET process between QDs and absorbers, Exciton ABS 642, a coordination compound obtained from Exciton Inc., was utilized as a quencher. The solutions of the QDs and the absorbers in toluene were prepared separately keeping an absorbance of 0.1 in their respective solutions. Subsequently, the absorber solution was added to the QDs to investigate the effect of the quencher on the fluorescence emission intensity and luminescence lifetime of the QDs. The resulting solution was found to be optically clear and homogeneous, and the final concentration of the QDs was maintained at an absorbance of ~ 0.05 for the time-dependent luminescence experiments.

## RESULTS AND DISCUSSION

The absorption spectrum of InGaP/ZnS core/shell QDs in toluene showed a clear excitonic absorption peak at 600 nm at room temperature (Fig.1), corresponding to the energy bandgap ($E_g$) of ~ 2.06 eV. The absorption spectrum shows a second peak at ~ 500 nm due to absorption by excited states of the The PL spectrum of the QDs showed a maximum at 650 nm with full width at half maximum of ~ 85 nm. The Stokes shift is found to be 50 nm (159 meV) for InGaP/ZnS core/shell QDs (Fig.1).

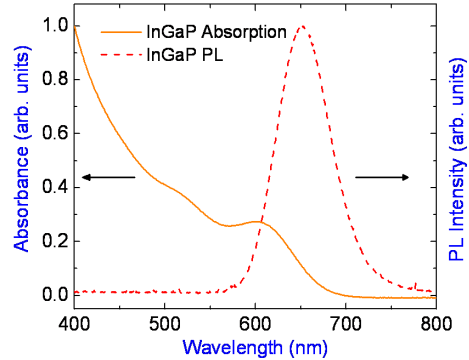

**Figure 1.** Optical absorption (solid line) and PL (dashed line) spectra of InGaP quantum dots.

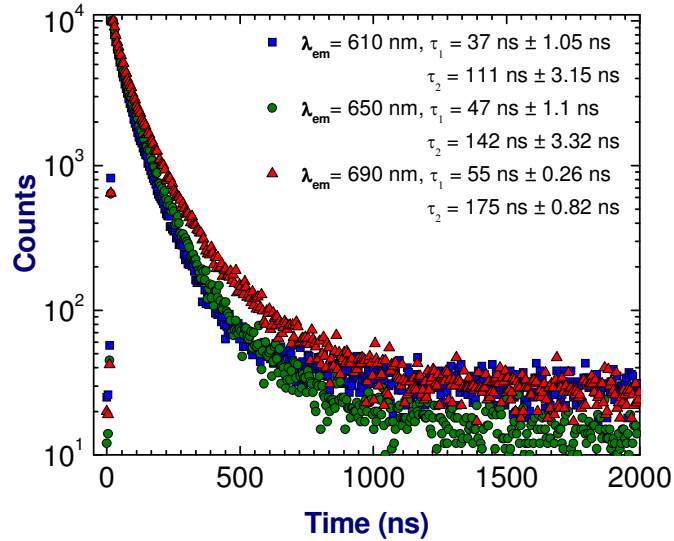

**Figure2**. Time resolved PL spectra of InGaP quantum dots at three different emission wavelengths 610 nm (blue squares), 650 nm (green circles) and 690 nm (red triangle), corresponding to the wavelengths below and above the emission maximum of 650 nm.

Time-resolved PL study of QDs provides information on the kinetics of carrier relaxation [15,18,19,20]. The PL decay of the QDs (Fig. 2) at the emission maximum (650 nm) followed a bi-exponential model ($\tau_1$= 47.1 ns (± 1.1ns), $\tau_2$ = 142 ns (± 3.32 ns)). The steady state PL measurements (Fig.1) exhibited a broad line-width (~ 85 nm), indicative of a large size-distribution in the QD ensemble. To verify this, we performed time-resolved PL measurements at emission wavelengths above (690 nm) and below (610 nm) the emission maximum (Fig. 2). The emission at the shorter wavelength, 610 nm, corresponds to smaller QDs possessing greater carrier confinement, and hence a larger oscillator strength resulting in a faster radiative decay [21]. Experimentally, we observed the lifetimes at 610 nm to be $\tau_1$= 36.5 ns (± 1.05 ns), $\tau_2$= 111 ns (± 3.15 ns). Conversely, the emission at a longer wavelength is expected to have its major contribution from larger QDs that have smaller oscillator strength causing a longer radiative lifetime [21]. We found the lifetimes of the QDs at 690 nm to be $\tau_1$= 55.7 ns (± 0.26 ns), $\tau_2$= 175 ns (± 0.82 ns).

Resonance energy transfer is an attractive optical method to probe distance-regulated and time-dependent biological events under both *in vivo* and *in vitro* conditions [10]. Real time spectroscopic ruler techniques, such as Förster resonance energy transfer (FRET), and Surface energy transfer (SET), fundamentally rely on the RET principle which is based on the kinetics of energy transfer process between a donor fluorophore and an acceptor molecule [3,10,13]. The efficiency of RET depends on the spectral overlap of a donor's luminescence with the absorption of a potential acceptor [10]. Considering the role of QD as a promising fluorescent donor, this study investigated the efficiency of the RET process between InGaP/ZnS core/shell QD and an absorber. Exciton ABS 642 exhibits an absorption peak at 650 nm, and has considerable overlap with the fluorescence emission of InGaP QDs (Fig. 3). Moreover, the absorber does not exhibit any fluorescence and scattering under the experimental conditions discussed here. In addition, the excitation source emits at 467 nm, which is far away from the absorption wavelength of the acceptor, minimizing the possibility of direct excitation of the acceptor molecule.

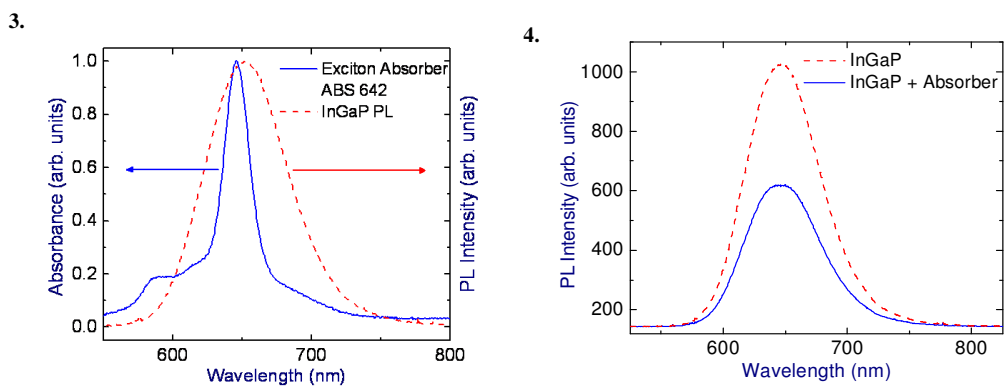

**Figure 3.** The absorption spectrum of the absorber ABS 642 showed a significant overlap with the photoluminescence spectrum of InGaP quantum dots at 650 nm. **Figure 4.** CW PL spectra exhibited the quenching of the InGaP QD luminescence. The emissions of the QD in the absence (dashed line) and in the presence (solid line) of the quencher (Exciton ABS 642) are shown.

Considerable quenching in the emission intensity of the InGaP/ZnS core/shell QDs was observed after mixing with Exciton ABS 642 (Fig. 4). Additionally, no shift in the original emission wavelength of the QDs was observed, which affirmed that the absorber molecules did not perturb the native chemical environment of the InGaP/ZnS core/shell QDs. The reduction of the emission intensity of the InGaP/ZnS core/shell QDs can be accounted for the energy transfer process between the QD and the absorber. Since the spatial proximity of a quencher molecule provides a new decay channel to dissipate energy of the photo-excited QD [3,10], the PL lifetime of the QD is expected to undergo an alteration due to the energy transfer process.

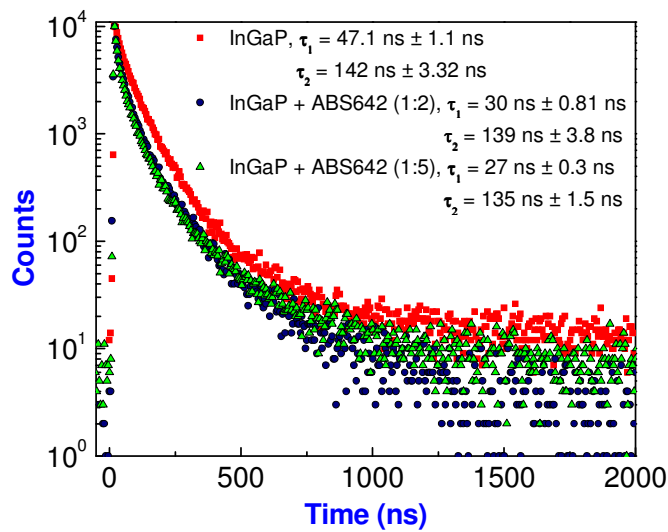

**Figure 5:** Time-resolved photoluminescence spectra of InGaP quantum dots in the presence of the quencher, Exciton absorber ABS 642. Additionally, the decay of the InGaP quantum dots (filled squares) is presented. All spectra were recorded at 650 nm.

Photoluminescence lifetime of InGaP/ZnS core/shell QDs was analyzed in the presence of the absorber to confirm this energy transfer process. The lifetime of the InGaP/ZnS core/shell QDs altered considerably in the presence of the absorbers (Fig. 5) at the emission maximum of 650 nm. The reduction of the first QDs' lifetime, $\tau_1$ (~ 36 % at 1:2 and 42 % at 1:5 QDs to absorber concentration ratio, respectively) in the presence of the absorber supports the inference that a RET process took place between the InGaP/ZnS core/shell QD and the absorber in the solution phase. Although there is no static link between the donor QDs and the acceptor molecules in the current experiment, the evidence of RET between the InGaP QDs and the ABS 642 absorber molecules indicates spatial proximity between them. Based on the aforesaid observations, we anticipate that biocompatible InGaP QDs can be applied towards investigating biological processes entailing longer time and relatively large distance changes [9], which are out of reach of conventional organic fluorophores. By reducing the effects of background autofluorescence [4,22], and enhancing the photo-stability of fluorescent labels, these QDs will help to elucidate relatively large (>10 nm) biomolecular conformational changes [9], such as DNA structural dynamics, protein folding, RNA folding, and oligomerization of membrane proteins [23,24].

**CONCLUSIONS**

In summary, the absorption and PL emission studies reported here show that InGaP QDs with a moderate stokes shift can be utilized as a promising fluorescent donor for spectroscopic ruler applications. The investigation of PL lifetime of InGaP/ZnS core/shell QDs depicted fluorescence decay with relatively longer lifetime compared with conventional bio-conjugating QDs (such as, CdSe, CdS). In addition, this study has demonstrated the possibility of using these

QDs as fluorescent donor in the RET mechanism. We anticipate the use of these less toxic InGaP colloidal QDs with longer luminescence lifetimes to further research in areas of biological imaging and biosensing.

## ACKNOWLEDGMENTS

This work was supported by the US Army Research Office (Grant no. W911 NF-07-1-0397). We thank Dr. I. Kuksovsky for useful discussions. We also acknowledge Evident Technologies for providing the InGaP quantum dots used in the present work. SC thanks the Graduate Center of CUNY for the Dissertation fellowship.